\documentclass[11pt,a4paper]{article}
\usepackage{cite,fullpage,graphicx}

\newcommand{\abc}[1]{\mbox{#1)}\quad}

\newcommand{\dd}{\mbox{d}}
\newcommand{\eqs}{Eqs.$\;$}
\newcommand{\eq}{Eq.$\;$}
\begin{document}

\title{Some cosmological consequences of the five-dimensional\\
Projective Unified Field Theory}

\author{
A.A. Blinkouski{\thanks{E-mail: blinkouski@yandex.ru}}\\
International Sakharov Environmental University\\
23 Dolgobrodskaya Str, Minsk 220009, Belarus \and
A.K. Gorbatsievich{\thanks {E-mail: gorbatsievich@bsu.by}}\\
Department of Theoretical Physics, Belarusian State University \\
4 Skaryna Ave., Minsk 220050, Belarus}

\date{}

\maketitle

\begin{abstract}
\noindent The classical observational cosmological tests (Hubble
diagram, count of sourc\-es, etc.) are considered for a homogeneous and
isotropic model of the Universe in the framework of the
five-dimensional Projective Unified Field Theory in which gravitation
is described by both space-time curvature and some hypothetical scalar
field ($\sigma$-field). It is shown that the presence of the
$\sigma$-field can essentially affect conclusions obtained from the
cosmological tests. The surface brightness-redshift relation can be
used as a critical test for $\sigma$-field effects. It seems reasonable
to say that the available experimental data testify that the
$\sigma$-field decreases with time. It is concluded that the spatial
curvature is positive or negative depending on whether the mass density
is larger or smaller than some critical parameter which is smaller than
the critical density and can even take negative values. It is shown
that the increase in the number of the observational cosmological
parameters as compared to the standard Friedmann model can essentially
facilitate coordination of the existing observational data.
\end{abstract}

\thispagestyle{empty}


\vspace{5ex}
\section{Introduction}

It has been known that the cosmological tests \cite {Sandage, Weinberg}
are a convenient method of studying cosmological gravitational fields.
The most important of them are: magni\-tude-redshift relation (Hubble
diagram), count of sourc\-es, angular size-redshift relation,  etc.
These tests allow one to find the Hubble constant~$H_{0}$ and the
deceleration parameter~$q_{0}$. However recent estimates of these
parameters, obtained from different tests in the framework of the
standard Friedmann model, are in rather poor mutual agreement without
special additional assumptions (see, e.g., \cite{Baryshev, Yoshii} and
references therein). The reasons for these difficulties can be both in
unreliability of the observational data (which is mainly connected with
evolution and selection effects) and in the restriction to the
Friedmann model based on the equations of General Relativity (GR). In
this context, a consideration of cosmological consequences of theories
generalizing GR deserves attention. One of such theories is the
5-dimensional projective unified field theory (PUFT) developed  by
E.~Schmutzer~\cite{Schmutzer68, Schmutzer83,Schmutzer95}.

As is well known, the idea of a 5-dimensional unified field theory goes
back to the works of Kaluza and Klein~\cite{Kaluza,Klein}. The pioneers
of the projective approach to this theory were Veblen and van
Dantzig~\cite{Veblen,Dantzig}. Later this approach was further
developed by many authors (the corresponding references and a review of
other higher-dimensional unified theories see
in~\cite{Vladimirov87,Vladimirov98,Overduin}).

In PUFT gravitation is described by both space-time curvature and some
hypothetical scalar field ($\sigma$-field). To characterize the scalar
field predicted in PUFT as a new fundamental phenomenon in Nature,
E.~Schmutzer introduced the notion ``scalarism'' (adjective:
``scalaric'') by analogy with electromagnetism. The source of this
``scalaric'' field can be both the electromagnetic field and a new
attribute of matter which Schmutzer has called ``scalaric mass''. It
should be noted that the presence of the $\sigma$-field can lead to
essential additions to the general picture of the Universe
evolution~\cite {Schmutzer1a, Schmutzer2a,Schmutzer2000}.

In this paper we shall consider a theory of classical cosmological
tests within the framework of PUFT. Also, we shall investigate the
observational parameters of a homogeneous and isotropic model on the
basis of PUFT. It~is obvious that the presence of the $\sigma$-field in
the theory leads to an extension of the number of the observational
cosmological parameters as compared to the standard Friedmann model.
This circumstance, from our point of view, will allow us to make
consistent the observational data existing now. That is primarily the
data of cosmological tests, the problem of dark matter, etc. (see
e.g.~\cite{Borner} and also~\cite{Baryshev, Yoshii}). All the results
obtained will be compared with similar predictions of the standard
Friedmann cosmology.

\section{Field equations of PUFT}

The version of PUFT investigated here is based on the  postulated
5-dimensional Einstein-like field equations. By projecting them into
the 4-dimensional space-time one  obtains the following 4-dimensional
field equations (the cosmological term is omitted
here)~\cite{Schmutzer83}:
\begin{eqnarray}\label{2.1}
R_{mn}-\frac{1}{2}\;g_{mn}\;R
=\kappa_0\left(E_{mn}+\Sigma_{mn}+\Theta_{mn}\right)
\end{eqnarray}
are the generalized gravitational field equations;
\begin{eqnarray}\label{2.2}
\abc{a} H^{mn}{}_{; n} = \frac{4\pi}{c}j ^ {m}, \qquad
\abc{b}B_{mn,k}+B_{km,n}+B_{nk,m} = 0, \qquad \abc{c} H_{mn} =
e^{3\sigma} B_{mn}
\end{eqnarray}
are the generalized electromagnetic field equations;
\begin{eqnarray} \label {2.3}
\sigma ^ {, k} {} _ {; k} = \kappa _ 0 \left (\frac {2} {3} \vartheta +
\frac {1} {8\pi} B_{ik} H^{ik} \right)
\end{eqnarray}
is the scalar field equation. Here $R_{mn}$ is the Ricci tensor,
\begin{eqnarray} \label{2.4}
E _{mn} = \frac {1} {4\pi} \left (B_{mk} H^{k}{}_{n} + \frac{1}{4}
g_{mn} B_{ik} H^{ik} \right)
\end{eqnarray}
is the electromagnetic energy-momentum tensor,
\begin{eqnarray}\label{2.5}
\Sigma_{mn} = - \frac {3} {2\kappa_0} \left (\sigma_{,m} \sigma_{,n} -
\frac{1}{2} g_{mn} \sigma _ {, k} \sigma ^ {, k} \right)
\end{eqnarray}
is the scalaric energy-momentum tensor, $\Theta_{mn}$~is the
energy-momentum tensor of the nongeometrized matter (substrate),
$H_{mn}$ and~$B_{mn}$ are the electromagnetic induction and the field
strength tensor, respectively, $j^{k}$~is the electric current density,
$\vartheta$~is the scalaric substrate density, $\kappa_{0}=8\pi
G/c^{4}$ is Einstein's gravitational constant ($G$~is Newton's
gravitational constant). Latin indices run from~1 to~4; the comma and
semicolon denote  partial and covariant derivatives, respectively; the
signature of the space-time metric is~$+2$.

These field equations lead to the following generalized energy-momentum
conservation law and continuity equation for electric current density:
\begin{eqnarray}\label {2.6}
\abc{a} \Theta ^{mn}{}_{;n} = -\frac{1}{c}B^{m}{}_{k}
         j^{k} + \vartheta \sigma^{,m} ,   \qquad
\abc{b} j^{m}{}_{;m} = 0.
\end{eqnarray}
Using (\ref {2.2}) and (\ref {2.6}) it is possible to show \cite
{Gorbatsievich96} that in PUFT, as well as in GR, light rays propagate
along null geodesics of space-time. However,
\begin{eqnarray}\label {2.8}
(e ^ {3 \sigma} T^{mn})_{;n}=0,
\end{eqnarray}
where $T^{mn}$ is the energy-momentum tensor of the photon beam. Thus
the scalar $\sigma$-field  can lead either to absorption of light or to
its amplification.

Concluding this section, it should be mentioned that E.~Schmut\-zer
since~1995 has preferred new non-Einstein-like 5-dimensional field
equations which  he applied to cosmology and cosmogony in a series of
papers~\cite{Schmutzer95,Schmutzer2000}. But this version of PUFT has
slightly different 4-dimensional field equations as compared with the
above-stated ones (one can find a detailed analysis of the geometric
axiomatics of PUFT in~\cite{Gorbatsievich2001}). It should be noted
that both variants are physically acceptable and deserve a
comprehensive study.

\section{Basic equations for a homogeneous and isotropic cosmological model}

Let us consider a homogeneous and isotropic cosmological model with the
Robertson-Walker line element in the well-known form:
\begin{eqnarray} \label{3.1}
\dd s^{2}=R^2(t) \left[ \frac {\dd r^2}{1-kr^2} + r^2 (\dd \theta^2 +
\sin^2 \theta \; \dd \varphi^2) \right] - c^2 \dd t^2,
\end{eqnarray}
where $R(t)$ is the scale factor and $k$ takes the values $0$ or $ \pm
1 $. For an electrically neutral continuum which is described by the
energy-momentum tensor of a perfect fluid the field equations
(\ref{2.1}) and (\ref{2.3}) in the metric (\ref{3.1}) lead  to the
following set of equations (the dot denotes a time derivative,
$\varrho$~is the mass density, $p$~is  the pressure):
\begin{eqnarray} \label{3.2f}
\frac{\ddot{R}}{R} = - \frac{\kappa_0 c^2}{6} \left( \varrho c^2  + 3p
\right) - \frac{1}{2} \dot{\sigma}^2 ,
\end{eqnarray}
\begin{eqnarray} \label{3.3f}
\frac{\ddot{R}}{R}  +  \frac{2(\dot{R}^2 + k c^2 )}{R^2} =
\frac{\kappa_0 c^2}{2} \left( \varrho c^2 - p \right),
\end{eqnarray}
\begin{eqnarray} \label{3.4f}
\ddot{ \sigma } + 3 \frac{  \dot{R}  }{R} \; \dot{  \sigma  }  = -
\frac{2}{3}\; \kappa_0 c^2 \vartheta ,
\end{eqnarray}
while the generalized energy conservation law (\ref{2.6}) gives
\begin{eqnarray} \label{3.5f}
\dot{\varrho} + 3 \frac{ \dot{R} }{R} \left(  \varrho  +  \frac{p}{c^2}
\right) = \frac{\vartheta}{c^2} \dot{\sigma} .
\end{eqnarray}
\eqs (\ref{3.2f}) to~(\ref{3.5f}) determine the  dynamics of the
cosmological model if the equations  of  state, i.e., $p=p(\varrho)$
and $\vartheta = \vartheta(\varrho)$, are known. The Friedmann model
corresponds to the special  case $\vartheta=0$ and $\dot{\sigma}=0$ of
our model. Unfortunately, the above set of differential equations
leads~\cite{Schmutzer1a} to an Abel equation and till now was solved
exactly only in some special cases~\cite{Schmutzer1a, Schmutzer2a,
Herlt, Blinkouski1,Blinkouski2}.

Now we examine light propagation in a Robert\-son-Walker space-time.
Consider light emitted from a point with the radial coordinate~$r_{1}$
at the time~$t_{1}$. The light, propagating along a null-geodesic line,
will be received at the point~$r=0$ and at the time~$t_{0}$ if
\begin{eqnarray} \label{3.2}
\int\limits_{t_1}^{t_0} \frac {c\; \dd t}{R(t)} =
\int\limits_0^{r_1}\frac{\dd r}{\sqrt{1-kr^2}}\;.
\end{eqnarray} Then the redshift of the light source is given by the usual
formula \begin{eqnarray} \label {3.3}
            1 + z = \frac {R(t_0)}{R(t_1)}\;.
\end{eqnarray}
On the other hand, the absolute luminosity~$L$ of the source and its
apparent bolometric luminosity~$\ell$ are connected by the relation
\begin{eqnarray}\label {3.4}
\ell(t_0)=\frac{L\; e^{-3 \Delta \sigma}}{4\pi(1 + z)^2 R^2 (t_0)\;
r^{2}_{1}}\; ,
\end{eqnarray}
where $\Delta\sigma\equiv\sigma(t_{0})-\sigma(t_{1})$. The presence of
the multiplier $e^{-3 \Delta \sigma}$ in the last expression is a
consequence of \eq(\ref{2.8}). Using~(\ref{3.4}), it is possible to
show that the flux density of radiation~$S(\nu)$ (i.e. the power per
unit area and per unit frequency interval of a receiver) is given by
\begin{eqnarray}\label {3.6}
    S (\nu) = \frac {P \left [\nu \left (1 + z \right) \right]}
    {(1 + z)R^{2}(t_0)\; r^{2}_{1}}\; e^{-3 \Delta \sigma} ,
\end{eqnarray}
where $P$ is the intrinsic source power per unit solid angle and per
unit frequency interval.

With (\ref{3.4}) the luminosity distance~$d_{\ell}$  to the source is
determined by the following expression:
\begin{eqnarray}\label{3.8}
     d_{\ell} \equiv \sqrt{\frac{L}{4\pi\ell }}
       = \left( 1 + z \right) r_{1} R(t_0)\; e^{ 3\Delta \sigma/2}.
\end{eqnarray}
If~$D$ is the linear size and~$\delta$ is the metric angular diameter
of the source, then the angular diameter distance~$d_{a}$ has the
form~\cite{Weinberg}:
\begin{eqnarray}\label{3.10}
    d_{a} \equiv \frac {D}{\delta}=\frac{R(t_0) r_1 }{1+z}\;.
\end{eqnarray}
From \eqs (\ref{3.8}) and (\ref {3.10}) we get
\begin{eqnarray}\label{3.11}
    d_{\ell} =  d_{a} ( 1 + z )^2 e^{\frac{3}{2} \Delta \sigma}.
\end{eqnarray}
Hence, taking into consideration the $\sigma$-field effects can  cause
changes in the construction of an extragalactic distance scale.

\section{Observational cosmological tests}

 In deriving theoretical relations that describe the cosmological tests we
 refer to the small~$z $ ($z\ll 1$) approximation. In this case we
need not integrate rather  complex cosmological equations of PUFT
(\ref{3.2f})--(\ref{3.5f}). It is only sufficient to require a relevant
regularity of the functions~$R(t)$ and~$\sigma(t)$.

\subsection{Hubble diagram}
First we consider the  $\sigma$-field influence on the relation
$d_{\ell}(z)$.  To this end, let us examine  sources with the same
intrinsic luminosity $L$. The quantities $d_{\ell}$ and $z$ of each
source are bound to its unknown coordinates by the relations (\ref
{3.2}), (\ref {3.3}) and (\ref {3.8}).  Assuming that $t-t_{0}$ and
$r_{1}$ are small ($t_{0}$ corresponds to the present epoch), we can
expand $R (t) $ and $e^{\sigma (t)} $ in the series
\begin{eqnarray}\label{4.1}
R(t)= R(t_0) \Bigl[1 + H_0 (t - t_0) - \frac{1}{2}\; q_0 H_0^2
(t-t_0)^2 + \cdots \Bigr],
\end{eqnarray}
\begin{eqnarray}\label{4.2}
e^{\sigma(t)} = e^{\sigma(t_0)} \left[1 + \lambda_0 H_0 (t-t_0) +
\cdots \right],
\end{eqnarray}
where $H_{0}$ and $q_{0}$ are the Hubble constant  and deceleration
parameter, respectively, defined  in the usual way:
\begin{eqnarray} \label{4.3}
H_0 \equiv \frac{ \dot{R}(t_0) }{ R(t_0) } \;, \qquad q_0 \equiv -
\frac{ \ddot{R}(t_0) R(t_0) }{ \dot{R}^2 (t_0) }\; .
\end{eqnarray}
The dimensionless parameter $\lambda _0$, characterizing the scalar
field, is given by
\begin{eqnarray}\label{4.4}
\lambda_0\equiv\frac{1}{H_0}\frac{\dd\sigma(t_0)}{\dd t}\;.
\end{eqnarray}
Taking into account \eqs (\ref{3.2}) and (\ref{3.3}), the expansions
(\ref{4.1}) and (\ref{4.2}) allow one to present $d_{\ell}$ (see
(\ref{3.8})) as a power series in $z$:
\begin{eqnarray}\label{4.8}
    d_{\ell} = \frac{c}{ H_0 }
    \Bigl[ z + \frac{1}{2} ( 1 - q_0 + 3\lambda_0) z^2 +\cdots \Bigr].
\end{eqnarray}
This relation can be rewritten as a formula for the apparent
luminosity:
\begin{eqnarray}\label{4.9}
    \ell = \frac{ L H_0^2 }{ 4 \pi c^2 z^2 }
            \Bigl [ 1 +(q_0 - 3\lambda_0 -1 ) z + \cdots \Bigr].
\end{eqnarray}
Thus, if the equations of PUFT are valid, then from (\ref{4.8}) and
(\ref{4.9}) it follows that, in the case $z\ll 1$, in astronomical
observations, some effective deceleration parameter
\begin{eqnarray}\label{4.10}
        \ q_0^{\rm{eff}} = q_0 - 3\lambda_0,
\end{eqnarray}
is measured, and the real deceleration parameter $q_{0}$ cannot be
obtained from the Hubble diagram.

\subsection{Counts of sources}

Let us assume that the number of sources per unit physical volume with
absolute luminosities within the bounds from $L $ up to $L+\dd L$ at
the time $t_{1}$ is $n(t_{1}, L)\; \dd L$. Then the number of sources
with radial coordinates from $r_{1} $ up to $r_{1} + \dd r_{1} $ is
given by
\begin{eqnarray}\label{4.11}
     \dd N = \frac{ 4\pi R^3 (t_1) r_{1}^2\; }
     {\sqrt{1 - k r_{1}^2}} \; n( t_1, L )\; \dd r_1 \;\dd L .
\end{eqnarray}
From this relation, by  taking account of (\ref{3.2}) we obtain that
the number of sources with redshifts smaller than $z $ and the apparent
luminosity greater than $ \ell $ is given by
\begin{eqnarray}\label{4.12}
   N ( {<}z,\ {>}\ell) = \int\limits_{0}^{\infty} \dd L
                \int\limits_{t_{m}}^{t_0}
                 c\; \dd t_1\; 4\pi r_{1}^2 R^2 (t_1)\; n( t_1, L ).
\end{eqnarray}
Here $t_m=\max \{t_z,t_{\ell}(L)\}$, where $t_z$ and $t_{\ell}(L)$ are
determined from \eqs (\ref{3.3}) and (\ref{3.4}), respectively:
\begin{eqnarray}\label{4.13a}
\abc{a} R( t_z ) = \frac{ R( t_0) }{ 1+z }\; , \qquad
 \abc{b} \frac{ r^2 (t_\ell)}{R^2 (t_\ell)}=
        \frac{L}{4\pi \ell R^4 (t_0)}\;
                e^{-3 [\sigma(t_0) - \sigma(t_\ell)]} .
\end{eqnarray}

As in Friedmann's cosmology \cite{Weinberg}, we shall assume that the
spectrum of all sources has the form $P \sim \nu ^ {-\alpha} $ with
$\alpha \approx 0.75 \;\!$. Then, using~(\ref {3.6}), we find that the
number of sources with redshifts smaller than~$z $ and with the flux
density at frequency $\nu$ greater than~$S $ is
\begin{eqnarray}\label{4.16}
    N ({<}z,\ {>}S;\ \nu ) =
\int\limits_{0}^{\infty} \dd P
          \int\limits_{t_{m}}^{t_0}
          c\; \dd t_1 \; 4\pi r_{1}^2 R^2 (t_1)\; n( t_1, P; \nu
          )\;.
\end{eqnarray}
Here $t_{m}=\max\{t_z,t_S (P)\}$, where $t_{S}(P)$ is a solution of the
equation
\begin{eqnarray}\label{4.17}
    r^2 (t_S) \left[ \frac{ R (t_S) }{ R (t_0) } \right]^{-1-\alpha} =
           \frac{ P(\nu)\; e^{ -3 [ \sigma(t_0) - \sigma(t_S) ] } }{ S(\nu)
                      R^2 (t_0) }\;,
\end{eqnarray}
and $n (t_1, P; \nu)\;\dd P$ is the space density  of sources with the
intrinsic power at frequency $\nu$ ranging from $P $ up to $P + \dd P
$.

In order to select $\sigma$-field effects, we shall restrict our
consideration to the case where there is no evolution of the sources.
This means that the sources are not born and do not disappear, and also
their luminosity does not depend on time. Then we have~\cite{Weinberg}:
\begin{eqnarray}\label{4.18}
    \abc{a} n( t, L ) = \left[ \frac{R(t_0)}{R(t)} \right]^{3} n( t_0, L
    )\;, \qquad
\abc{b} n( t, P; \nu ) = \left[ \frac{R(t_0)}{R(t)}
     \right]^{3} n( t_0, P; \nu )\;.
\end{eqnarray}
At low $z$ and large $\ell$ or $S$ we can use the
expansions~(\ref{4.1}) and~(\ref{4.2}). In this case, from \eqs
(\ref{4.12})--(\ref{4.18}) we find
\begin{eqnarray}\label{4.20}
N({<}z) = \frac{4\pi}{3} \frac{c^3}{H_0^3}\; z^3
           \int\limits_{0}^{\infty} \dd L \;n( t_0, L )
           \left[ 1 - \frac{3}{2} ( 1 + q_0 ) z + \cdots  \right],
\end{eqnarray}
\begin{eqnarray}\label{4.21}
N({>}\ell ) = \frac{4\pi}{3}\; (4\pi \ell)^{-3/2}
\int\limits_{0}^{\infty} \dd L \; n( t_0, L ) L^{3/2}
     \left[ 1 -  3\Bigl( 1 + \frac{3}{2} \lambda_0 \Bigr)
         \frac{H_0}{c} \left( \frac{L}{4\pi \ell}
         \right)^{1/2} + \cdots \right],
\end{eqnarray}
\begin{eqnarray}\label{4.22}
N({>}S, \nu ) = \frac{4\pi}{3}\; S^{-3/2}
         \int\limits_{0}^{\infty} \dd P \; n( t_0, P; \nu ) P^{3/2}
         \left[ 1 - \frac{3}{2} \left( 1 + \alpha + 3\lambda_0 \right)
         \frac{H_0}{c} \left( \frac{P}{S} \right)^{1/2} + \cdots \right].
\end{eqnarray}

Notice that \eq (\ref{4.20}) coincides with a similar result of GR.
Thus at~$z\ll 1$ the $\sigma$-field does not affect the magnitude $N(<
z)$. It is obvious that this follows  from~(\ref{4.13a}). Hence, in
principle, the experimental values $N(< z)$ at low~$z$ could be used
for determining the real deceleration parameter~$q_{0}$. At the same
time, measurements of $N(>\ell)$ or $N(>S, \nu)$ at large~$\ell$ or~$S$
do not give any information about~$q_{0}$. But these measurements could
be used to determine the parameter~$\lambda_{0}$.

However, it should be noted that, as well as in the standard Friedmann
cosmology, \eqs (\ref{4.20}), (\ref{4.21}) and (\ref{4.22}) are in
conflict with observational data (see, e.g., \cite{Weinberg, Borner}
and references therein). For example~\cite{Weinberg}, the counts of
radio sources testify that the function $N(> S,\;\nu)$ decreases with
growing~$S$ (at $S>5{\cdot}10^{-26}\;{\mbox{W}}{\cdot}{\mbox{\rm
m}}^{-2}{\mbox{Hz}}^{-1}$) approximately as $S^{-1{.}8}$ and definitely
faster than $S^{-3/2}$, and only at low $S$ it begins to decrease
slower than $S^{-3/2}$. Notice that according to (\ref{4.22}) the
function $N(> S,\nu)$ will decrease as $S^{-3/2}$ or faster provided
that $\lambda_{0} < -(1 + \alpha)/3 \approx - 0{.}58\,$. However, it is
difficult to explain such a complicated behaviour of the empirical
function $N(> S,\nu)$ only by means of the $\sigma$-field effects.
Consequently, it is necessary to take into account the evolution of the
sources. But in this case the reliability of the results obtained
depend on the reliability of evolutionary suppositions. Under this
circumstance the determination of cosmological parameters by means of
the above test, including the parameter $\lambda_{0}$, becomes very
complex.

\subsection{Angular size-redshift relation}

At low redshifts \eq (\ref{3.10}) for $d_{a}$, taking account of
(\ref{4.1}) and (\ref{4.2}), can be rewritten as
\begin{eqnarray}\label{4.24}
    d_{a} \equiv \frac {D}{\delta} = \frac{cz}{ H_0 }
        \left[ 1 - \frac{3}{2} \Bigl( 1+\frac{q_{_0}}{3}
                \Bigr) z + \cdots \right].
\end{eqnarray}
This outcome completely coincides with the similar result of standard
cosmology. Hence, the $\sigma$-field does not influence the relation
$\delta (z)$ at $z\ll 1$. Unfortunately, we cannot determine $q_{0}$
from this relation, because at low redshifts observational errors are
much greater than the differences in $q_{0}$ expected for different
cosmological models \cite{Baryshev,Kellermann}.

It is well to bear in mind that, generally speaking, at high redshifts
the function $\delta (z)$ will depend  on the parameter $\lambda_{0}$,
because the $\sigma$-field is present implicitly in $d_a$ according to
(\ref{3.10}). But $R(t)$ and~$r(t)$, contained in (\ref{3.10}), depend
on the $\sigma$-field. It is evident that this remark is correct for
the test $N(< z)$ at high redshifts too.

\subsection{Surface brightness-redshift relation}

From \eqs (\ref{3.4}) and (\ref{3.10}) it follows that the observed
surface brightness of sources is given by
\begin{eqnarray}\label{4.25}
 B \equiv \frac {\ell}{\delta^2} =
           \frac{L}{4\pi D^2 } \frac{e^{ -3 \triangle \sigma}}{(1+z)^4} =
           \frac{\mbox{const}}{(1+z)^4}\; e^{ -3 \triangle \sigma}\; ,
\end{eqnarray}
where we assume that all these sources are identical, i.e.
$\displaystyle L/D^2 = \mbox{const}$. Thus the presence of the
$\sigma$-field can essentially change the simple surface
brightness-redshift relation arising within the framework of~GR:
$$
  B =  \frac{\mbox{const}}{(1+z)^4}  \qquad \mbox{(GR)}.
$$
In the work \cite{Hubble} this equation was proposed to be used as a
test for the redshift nature. Obviously in the framework of PUFT \eq
(\ref{4.25}) can be used as a test for the presence of cosmological
$\sigma$-field effects. From~(\ref{4.25}), taking account
of~(\ref{4.1}) and~(\ref{4.2}),  we find
\begin{eqnarray}\label{4.27}
   B(z ) = \frac {L}{4\pi D^2}
   \left[\vphantom{\frac{L}{D^2}} 1 - ( 4 + 3\lambda_0 ) z + \cdots \right].
\end{eqnarray}
Notice that the parameters $q_{0}$ and~$H_{0}$ are absent in this
expression. Consequently, at low redshifts the surface
brightness-redshift relation allows one to estimate the $\sigma$-field
effects in a pure form.

In the work  \cite{Sandage91}, the relation $B(z)$ for a family of
giant elliptical galaxies with small~$z$ was investigated within the
framework of the standard Friedmann model. In this paper the observed
curve for the dependence of~$B$ on $\log (1+z)$ is just a little more
slanting than the straight line with a slope equal to~$-4$. According
to~(\ref{4.27}), it means that the parameter $\lambda_{0} < 0$ if only
we neglect the evolution effects. Hence, the $\sigma$-field has to
decrease with time if we assume that~$\sigma (t)$ is a monotonic
function.


\section{Cosmological parameters}

\subsection{Mass density and spatial curvature}

It should be noted  that for a correct interpretation of observational
data in the framework of PUFT, obtained, in particular, from the
cosmological tests, it is necessary to establish primarily a
relationship between the observational cosmological parameters of PUFT
and the mass density and spatial curvature of the present Universe. In
order to solve this problem, we need the cosmological equations of PUFT
(\ref{3.2f})--(\ref{3.5f}). From \eqs (\ref{3.2f}) and~(\ref{3.3f})
with~(\ref{4.3}) and~(\ref{4.4}) one can find
\begin{eqnarray} \label{b31}
\varrho_0 = \frac{3}{ \kappa_0 c^4 } \left[ \frac{ k c^2 }{ R_0^2 }
   + H_0^2 \Bigl( 1 - \frac{ \lambda_0^2 }{4} \Bigr) \right] ,
\end{eqnarray}
\begin{eqnarray} \label{b32}
p_0 = - \frac{1}{ \kappa_0 c^2 } \left[\frac{ k c^2 }{ R_0^2 } + H_0^2
\Bigl(  1 - 2 q_0 + \frac{3}{4} \lambda_0^2 \Bigr)\right] .
\end{eqnarray}
From the latter  equation we obtain that the spatial curvature~$k/R^2$
is positive or negative depending on whether the mass density is larger
or smaller than some critical parameter~$\chi_{c}$:
\begin{eqnarray} \label{b35}
\chi_c  = \varrho_c ( 1 - \lambda_0^{2}/{4}),
\end{eqnarray}
where $\varrho_c =  3 H_0^2/(\kappa_0  c^4)$ is the so-called critical
density. Thus in PUFT the type of the Universe (open, spatially flat or
closed) results from the  comparison of~$\varrho_{0}$ with~$\chi_{c}$
instead of the comparison~$\varrho_{0}$ with~$\varrho_{c}$. Notice that
the parameter~$\chi_{c}$ takes negative values if $|\lambda_{0}|>2$.

It is convenient to introduce the dimensionless density parameter by
\begin{eqnarray}\label{b36}
    \Omega_0 \equiv \varrho_0 /\!\varrho_c
\end{eqnarray}
and the dimensionless critical parameter by
\begin{eqnarray}\label{b37}
    \Omega_c \equiv \chi_c /\! \varrho_c  =1 - \lambda_0^2/4 .
\end{eqnarray}
This equality results in that $\Omega_c < 1$ is only valid if
$\lambda_{0}\neq 0$. Notice that the Universe is closed if $\Omega_0 >
\Omega_c$ and it is open if $\Omega_0 \leq\Omega_c$.

Let us now find out how the spatial curvature and the mass density
$\varrho_{0}$ or~$\Omega_{0}$ are connected with the observational
cosmological parameters of PUFT $q_0 $, $H_0 $ and~$\lambda_{0}$. In
the case of the dust model ($p=0$), from \eqs (\ref{b31}), (\ref{b32})
and~(\ref {b36}) we find
\begin{eqnarray} \label{b39}
\frac{k c^2}{R_0^2} = H_0^2 \left(2 q_0 - \frac{3}{4}\lambda_0^2 -
1\right),
\end{eqnarray}
\begin{eqnarray} \label{b310}
     \varrho_0 = \varrho_c ( 2 q_0 - \lambda_0^2),
     \qquad
     \Omega_0 = 2 q_0 - \lambda_0^2 .
\end{eqnarray}
Taking into account (\ref{b39}), one can obtain the conditions
determining the type of the Universe:
\begin{eqnarray} \label{b312}
\begin{array}{llll} q_0 - \displaystyle\frac{3}{8} \lambda_0^2 > \frac{1}{2}
         & \Rightarrow & k = +1  & \left( \Omega_0 > \Omega_c \right) ,
\medskip \\
 q_0 - \displaystyle\frac{3}{8} \lambda_0^2 = \frac{1}{2}
         & \Rightarrow &  k  =  0  &\left( \Omega_0 = \Omega_c \right) ,
\medskip\\
 q_0 - \displaystyle\frac{3}{8} \lambda_0^2 < \frac{1}{2}
         & \Rightarrow & k = -1  & \left( \Omega_0 < \Omega_c \right).   \\
\end{array}
\end{eqnarray}
In the case of a radiation-dominated Universe ($p=\varrho c^2/3$) \eqs
(\ref{b39})--(\ref{b312}) have the form
\begin{eqnarray} \label{b313}
\frac{ k c^2 }{ R_0^2 }  =  H_0^2  \left( q_0  -  \frac{1}{4}
\lambda_0^2 - 1 \right),
\end{eqnarray}
\begin{eqnarray} \label{b314}
\varrho_0 = \varrho_c  ( q_0 - \frac{1}{2} \lambda_0^2 ), \qquad
\Omega_0 = q_0 - \frac{1}{2} \lambda_0^2,
\end{eqnarray}
\begin{eqnarray} \label{b316}
\begin{array}{llll}
 q_0 - \displaystyle\frac{1}{4} \lambda_0^2 > 1
         & \Rightarrow & k = +1  & \left( \Omega_0 > \Omega_c \right) ,
\medskip\\
 q_0 - \displaystyle\frac{1}{4} \lambda_0^2 = 1
         & \Rightarrow &  k  =  0  &\left( \Omega_0 = \Omega_c \right) ,
\medskip \\
 q_0 - \displaystyle\frac{1}{4} \lambda_0^2 < 1
         & \Rightarrow & k = -1  & \left( \Omega_0 < \Omega_c \right).   \\
\end{array}
\end{eqnarray}
Thus in PUFT, unlike to  the Friedmann's cosmology, by  measuring only
the deceleration parameter~$q_{0}$ it is impossible to determine
whether the Universe is closed or open. For this purpose it is
necessary to have the values of the two parameters, $q_0$
and~$\lambda_0$ or $\Omega_0$ and~$\lambda _0$.

\subsection{Admitted regions for parameters}

It is interesting to note that in PUFT a spatially flat Universe can be
realized for the whole range of values of the mass
density~$\varrho_{0}$,
\begin{eqnarray} \label{b318}
0 \leq \varrho_0 \leq \varrho_c \;,
\end{eqnarray}
because $\varrho_0 = \varrho_c (1 -\lambda_0^2 /4)$ if $k =  0$.
However, the condition~(\ref{b318}) is necessary but not sufficient for
the 3-dimensional space to be flat. Recall that in the Friedmann model
the Universe is flat if and only if $\varrho_{0}=\varrho_{c}$. Taking
into account this circumstance, it is useful to study in more detail
the parameters of the theory.

First of all, let us find physically admitted regions for the
parameters $q_0$ and $\lambda_0$. To this end we shall rewrite the
natural inequality $\varrho_0\geq 0$ taking into account~(\ref{b310})
and~(\ref{b314}):
\begin{eqnarray}\label{b43}
 q_0 \geq \lambda_0^{2}/2 \;.
\end{eqnarray}
It is just the inequality which determines the admitted region of the
parameter~$q_0$ depending on~$\lambda_0$ (this region  is shaded in
Fig.~1). Note that~(\ref{b43}) is valid for both cases  $p=0$ and
$p=\varrho c^2 /3 $. For~$p = 0$, using~(\ref{b312}) and~(\ref{b43}),
we obtain (see Fig.~1) that if $q_0>2$ or $|\lambda _{0}|>2$, then a
closed Universe is only possible ($k=+1$), and if $0\leq q_0<1/2$, then
an open Universe is only possible ($k=-1$), while for $1/2\leq q_0 \leq
2$ all three types of the Universe are possible depending on the value
of $ \lambda_{0}$.

\vspace{1ex}
\begin{figure}\begin{center}
  \includegraphics[width=0.7\textwidth]{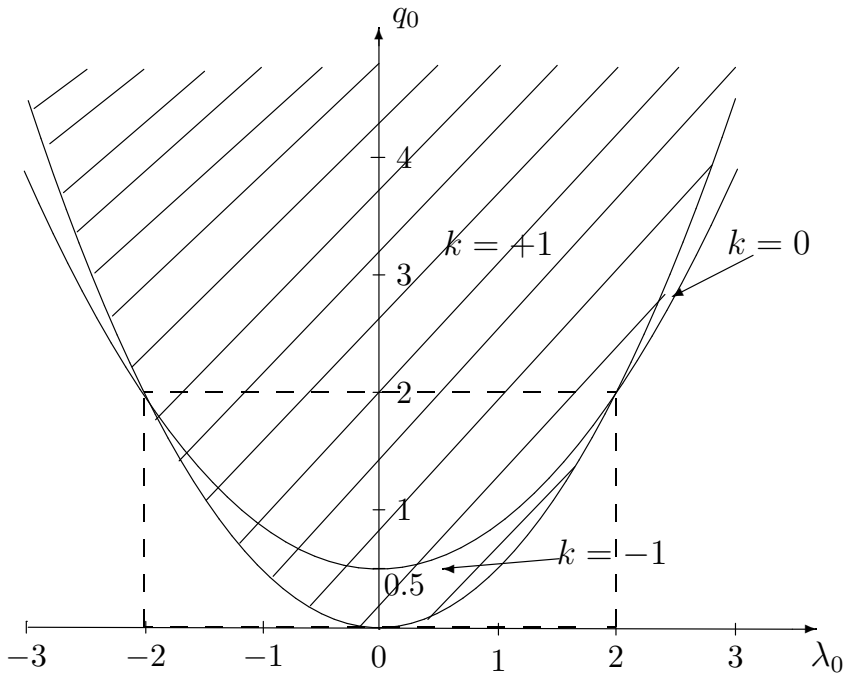}\\
  \caption{The admitted region of the parameter $q_0$ depending on $ \lambda _0 $.
The parabola corresponding $k=0$ is given for the case of the dust
model ($p=0 $).}\label{fig1}\end{center}
\end{figure}

Let us now take into consideration the available experimental data on
the magnitude-redshift relation. These data, obtained in the framework
of standard Friedmann cosmology (see, e.g., \cite{Baryshev, Yoshii}),
lead to the estimate $0{.}5 \leq q_0\leq 1$. In the case of PUFT,
taking into account \eq (\ref{4.10}) and the remarks about this
equation, we can suppose that the estimate $0{.}5 \leq q_0^{\rm
eff}\leq 1$ is sufficiently reliable. On this basis we shall determine
the possible values of the parameters~$q_0$ and~$\lambda_0$. Above all
we note that the inequality~(\ref{b43}) is consistent with \eq
(\ref{4.10}) if $q_0^{\rm eff} \geq -4{.}5\;\!$. From~(\ref{b43})
and~(\ref{4.10}) we find
\begin{eqnarray}
    \label{b54}  \begin{array}{ccc}  q_0^{\rm eff}  =  0{.}5  \quad   &
    \Longrightarrow & \left\{ \begin{array}{l}
     0{.}013 \leq q_0 \leq 19 \;  ,                    \\
     -0{.}16 \leq \lambda_0  \leq 6{.}2 \;  ,          \\
\end{array}
\right.\\
\vdots & \vdots &  \vdots\\
 q_0^{\rm eff} = 1{.}0 \quad  & \Longrightarrow &
     \left\{
\begin{array}{l}
0{.}05 \leq q_0 \leq 20 \;  ,                    \\
-0{.}32 \leq \lambda_0  \leq 6{.}3 \; .
\end{array} \right.
\end{array}
\end{eqnarray}
From (\ref {b54}) we learn  that, at large $q_{0}^{\rm eff}$, models
with the parameter~$q_0$ tending to zero are possible. This
circumstance can be used for coordination of experimental data of the
different tests. It should be noted that in the framework of the
Friedmann's cosmology (see e.g. \cite{Baryshev, Yoshii}) the very low
values for the deceleration parameter~$q_0$, obtained from counts of
sources, contradict  the above-mentioned values of~$q_0$ which follow
from the magnitude-redshift relation.

\section{Conclusions}

We have considered the classical cosmological tests (Hubble diagram,
count of sources, etc.) for a~homogeneous and isotropic model of the
Universe in the framework of the 5-dimensional Projective Unified Field
Theory. The results show that the presence of the scalar $\sigma
$-field predicted by PUFT can essentially affect the conclusions
obtained from the cosmological tests. We have shown, in particular,
that in PUFT the deceleration parameter~$q_{0}$ cannot be found from
the Hubble diagram at low redshifts. We can only measure  some
effective deceleration parameter~$q_0^{\rm eff}$ given by (\ref{4.10}).
It should be noted that  all the expressions describing cosmological
tests in the small~$z$ approximation do not depend on the choice of a
specific model (the spatial curvature sign, choice of the equation of
state, etc.). The  surface brightness-redshift relation can be used as
a critical test for $\sigma $-field effects, because the $\sigma$-field
can essentially change the simple dependence of the surface brightness
on the redshift in the form $B\sim {(1+z)}^{-4}$ which results from
Friedmann's cosmology. It seems reasonable to say that the available
experimental data testify that the $\sigma$-field decreases with time.

It is interesting to note that in cosmology, on the basis of the
version of PUFT investigated here, the spatial curvature is positive or
negative depending on whether the mass density is larger or smaller
than some critical parameter~$\chi_{c}$ determined by (\ref{b35}).
Moreover, the parameter~$\chi_{c}$ is smaller than the critical density
and can even take negative values. It should be emphasized that we did
not take into account a cosmological constant in the field equations.
On such a basis, in PUFT, a flat Universe with the current density
parameter $\Omega_{0}< 1$ is possible. These results can be used for
solving the dark matter problem. Recall that in Friedmann's cosmology
the inflationary prediction of flat Universe is at odds with the
current determinations of the matter density. Also, in PUFT the
increase in the number of observational cosmological parameters in
comparison with the standard Friedmann model can essentially facilitate
the co-ordination of the observational data existing now. However, the
comparison of cosmological theory with observations becomes technically
more complicated.

\section*{Acknowledgements}

The authors would like to thank Prof. Ernst Schmutzer for helpful
discussions and valuable remarks.


\end{document}